\documentclass{PoS}
\usepackage{graphicx}
\usepackage{epsfig}
\usepackage{rotating}

\def\bra#1#2{\ifx#2\ket\langle#1\else\langle#1\vert\fi#2}
\def\ket#1{\vert#1\rangle}

\leftline{\parbox{3cm}{\large\rm  DESY 07-179\\HU-EP-07/51\\LMU-ASC 56/07}
\vspace{1cm}}

\title{Localization of overlap modes and topological charge, 
vortices and monopoles in $SU(3)$ LGT}

\ShortTitle{Overlap modes, topological charge,
vortices and monopoles}

\author{\speaker{Ernst-Michael~Ilgenfritz}$^{\hspace{1mm}a}$, Karl~Koller$^{b}$, Yoshiaki~Koma$^{c}$, Gerrit~Schierholz$^{d,e}$, Thomas~Streuer$^{f}$, Volker Weinberg$^{d,g}$ and Markus~Quandt$^{h}$\\
~\\

$^{a}$ Institut f\"ur Physik, Humboldt Universit\"at zu Berlin, 12489 Berlin, Germany\\
$^{b}$ Sektion Physik, Universit\"at M\"unchen, 80333 M\"unchen, Germany\\
$^{c}$ Numazu National College of Technology, Numazu 410-8501, Japan\\
$^{d}$ John von Neumann-Institut f\"ur Computing NIC, 15738 Zeuthen, Germany\\
$^{e}$ Deutsches Elektronen-Synchrotron DESY, 22603 Hamburg, Germany\\
$^{f}$ Dep. of Physics and Astronomy, University of Kentucky, Lexington, KY 40506-0055, USA \\
$^{g}$ Institut f\"ur theoretische Physik, Freie Universit\"at Berlin, 14196 Berlin, Germany\\
$^{h}$ Institut f\"ur theoretische Physik, Universit\"at T\"ubingen, 72076 T\"ubingen, Germany\\ 
E-mail: \email{ilgenfri@physik.hu-berlin.de} 
}

\abstract{We present selected recent results of the QCDSF collaboration on the 
localization and dimensionality of low overlap eigenmodes and of the topological 
density in the quenched $SU(3)$ vacuum. We discuss the correlations between the 
topological structure revealed by overlap fermions without filtering
and the confining monopole and P-vortex structure obtained in the 
Indirect Maximal Center Gauge.}

\FullConference{The XXV International Symposium on Lattice Field Theory\\
                 July 30 - August 4, 2007\\
                 Regensburg, Germany}

\begin{document}

\section{Motivation}
\vspace{-0.2cm}
Overlap fermions~\cite{Neuberger} possess exact chiral symmetry on the lattice, 
realize the Atiyah-Singer index theorem~\cite{Hasenfratz} and provide a local 
definition of the topological charge density~\cite{Niedermayer}.
This makes them an ideal tool for investigating the chiral and topological QCD vacuum
structure.

In this talk we shall summarize some results of a recent extended study~\cite{Ilgenfritz1} 
of the vacuum structure of quenched QCD at zero temperature. One of the lessons  
is that there exists a whole family of topological descriptions, ranging 
from an UV filtered density (characterized by a cut-off scale $\lambda_{\rm cut}$ and 
sign-coherent selfdual clusters) to a topological density with high $O(a)$ resolution 
(also called ``all-scale density'') forming global, sign-coherent lower-dimensional 
structures~\cite{Horvath}. 
The other lesson is that, apart from the overall chirality that clearly distinguishes
zero and non-zero modes, the localization features are smoothly changing from 
one to the others. The non-zero modes, through the local chirality, still feel 
the (filtered) topological background.

This talk focuses on two aspects of Ref.~\cite{Ilgenfritz1}, the localization of
the low-lying eigenmodes and of the all-scale topological density.
The localization properties of the low-lying modes~\cite{Aubin,Gubarev,Koma,deForcrand} 
have attracted interest since they are hypothetically pinned down on singular 
defects~\cite{Zakharov} which are responsible for confinement. Candidates for this
role are monopoles and 
vortices as located by Abelian or center projection (for a review see~\cite{Greensite}). 
The localization of the all-scale topological density has also been 
considered~\cite{Aubin} 
for similar reasons. In particular, peaks are expected on vortex intersections 
etc. and usually searched for by zero modes~\cite{Reinhardt,Solbrig}. 
We stress that the mechanism behind the formation of the peculiar singular and 
global structure~\cite{Horvath} appearing at low density is unknown. This structure is 
necessary for the negativity~\cite{negativity,Koma,Ilgenfritz1} of the two-point 
function $C(x-y)=\langle q(x) q(y) \rangle$  
required by reflection positivity~\cite{Seiler}. This aspect of topological charge 
is complementary to the instanton-like clustering of the UV filtered density in
approximately (anti-)selfdual domains~\cite{Ilgenfritz1}.
Thus, the low-lying modes and the (unfiltered) topological density can be seen in 
closer relation to the mechanism of confinement. On a macroscopical level, 
such a relation is well established: the removal of vortices or monopoles from lattice 
configurations simultaneously destroys the topological charge and restores chiral 
symmetry~\cite{dElia,Boyko,Bornyakov}.

\section{Localization of overlap eigenmodes}
\begin{figure}[t]
\centering
\includegraphics[width=9cm]{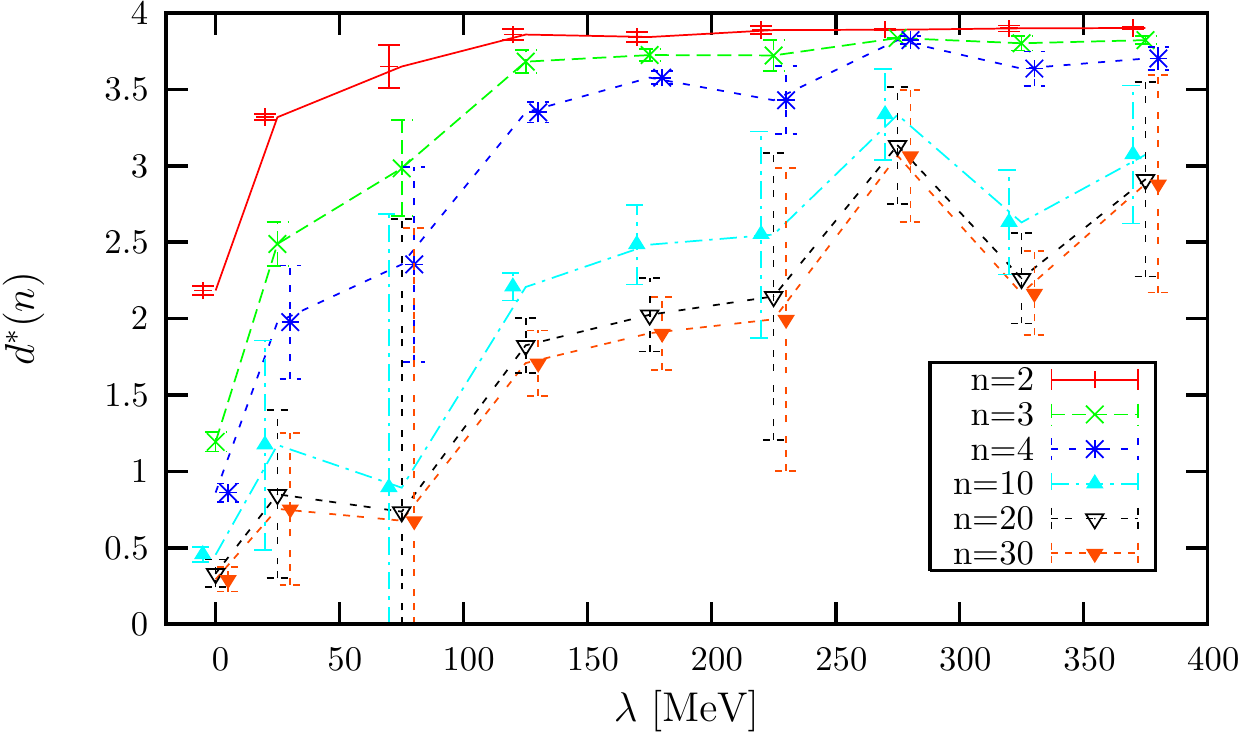}
\vspace{-0.4cm}
\caption{Fractal dimension $d^{*}(n)$ obtained from fits of
the volume dependence of the averages of $I_n$, for zero modes and for 
non-zero modes in bins $\Delta \lambda = 50 {\rm~MeV}$, for three 
ensembles with the same $\beta=8.45$ and different volumes.}
\label{fig:dimensions_from_Ip}
\end{figure}

\vspace{-0.2cm}
We use the massless Neuberger~\cite{Neuberger} overlap Dirac operator
\begin{equation}
D_{ov}(0)=\frac{\rho}{a}\left(1 + D_{W}/\sqrt{D_{W}^{\dagger}D_{W}} \right) \; , 
\quad {\rm with~~} D_{W} = M - \frac{\rho}{a} \; ,
\end{equation}
the Wilson Dirac operator with hopping term $M$ and negative mass $\rho/a$. 
The quenched ensembles of~\cite{Ilgenfritz1}
were generated by the L\"uscher-Weisz action, for $\beta=8.45$  
on lattices $12^3\times24$, $16^3\times32$ and $24^3\times48$, 
for $\beta=8.1$ on $12^3\times24$ and for $\beta=8.0$ on $16^3\times32$. 
First results have been reported by 
Y. Koma~\cite{Koma} at Lattice 2005. At that time, we estimated the dimension of 
zero modes and non-zero modes in the lowest bins of the spectrum from the volume 
$V$ dependence of the Inverse Participation Ratios (IPR), 
$IPR = V I_2 = V \sum_x |\psi_{\lambda}(x)|^4$,
averaged over the respective modes. The average IPR should follow a power law
\begin{equation}
\langle IPR \rangle = c_1 + c_2 V^{1-d^{*}/4} \; , 
\end{equation}
allowing to infer the fractal dimension $d^{*}$. We concluded~\cite{Koma} 
that zero modes are $d^{*}=2$ and next-to-zero modes $d^{*}=3$ dimensional.  
Now we refine this statement by considering generalized IPR's~\cite{Kravtsov}.
With their help one should be able to find lower dimensions for regions of higher 
density if a multifractal structure is physically realized. 
The second moment $I_2$ of the scalar density $p(x)=|\psi_{\lambda}(x)|^2$ is 
replaced by higher one, $I_n=\sum_x |\psi_{\lambda}(x)|^{2n}$, such that a 
sequence of dimensions $d^{*}(n)$ can be extracted from the volume scaling of 
$\langle I_n \rangle \propto L^{d^{*}(n)\left(n-1\right)}$. 
The result of this analysis is shown in Fig.~\ref{fig:dimensions_from_Ip}.
This plot shows that the regions of higher scalar density are lower dimensional 
(between $d^{*}=0$ and 1). There is a gradual change of the localization properties 
from zero modes to non-zero modes.

In Ref.~\cite{Ilgenfritz1} we have described methods to estimate the
dimension of an arbitrary distribution at {\it any} level of the density.
Both methods assume a cluster analysis already made to separate peaks 
of the distribution from the rest of the system. The emerging set of connected 
clusters, as function of a running parameter, {\it e.g.} the lower density cut-off for
the clusters, characterizes the distribution~\cite{Ilgenfritz1}. 
In the {\it random walker method}, for random walkers moving inside a cluster, 
the return probability to the cluster center, $P(0,t) \propto t^{-d^{*}/2}$, 
provides an estimate of the dimension $d^{*}$ depending on the adopted cut-off. 
Another cut-off dependent dimension $d^{*}$ can be inferred, in the {\it covering-sphere 
method}, from the growth (from 0 to 1) of 
the cumulative fraction $Q_{\rm cumulative} $ of a cluster's total charge that is 
covered by a 4D sphere of radius $R$. 
This growth begins $Q_{\rm cumulative} \propto R^{d^{*}}$~\cite{Ilgenfritz1}~\footnote{Although 
the results are similar, the effective dimensions obtained by the two methods do not 
strictly agree.}.

\begin{figure}[t]
\centering
\begin{tabular}{cc}
\includegraphics[height=4.5cm,width=7.1cm]{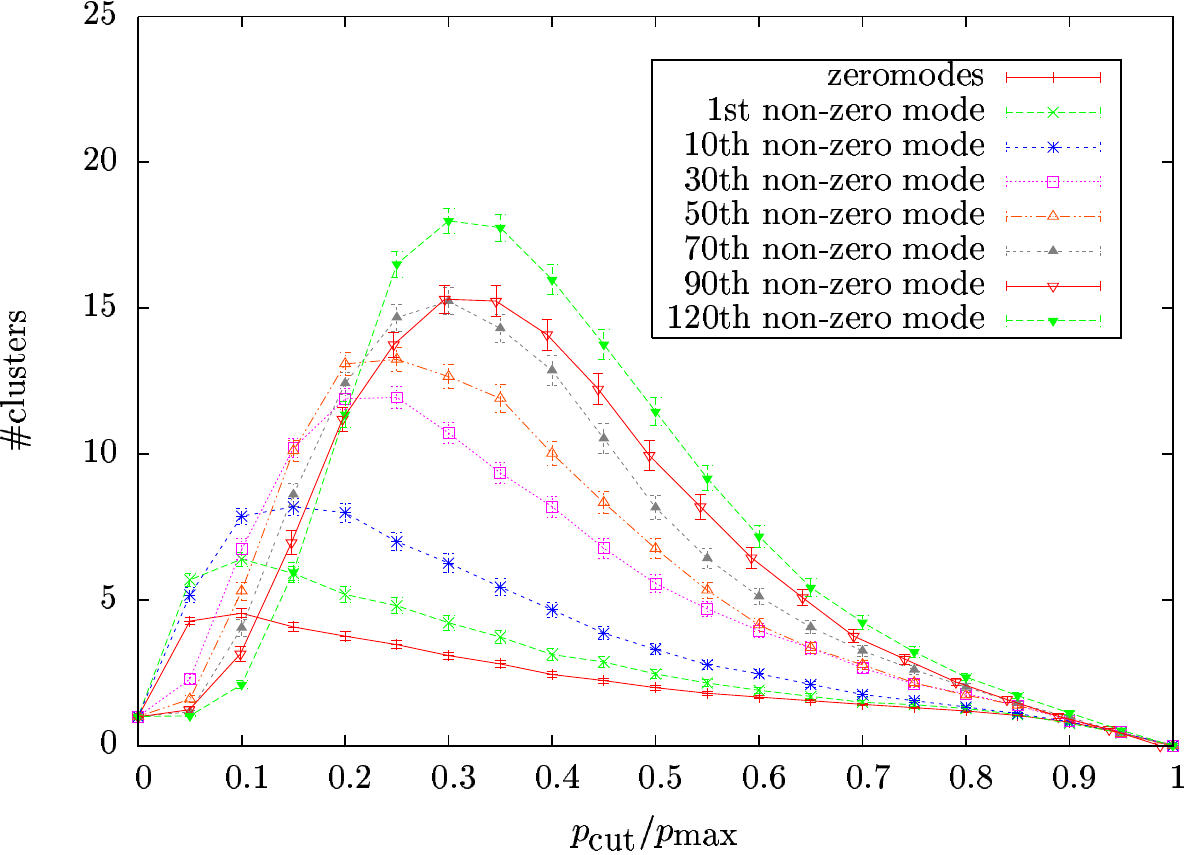}&
\includegraphics[height=4.5cm,width=7.1cm]{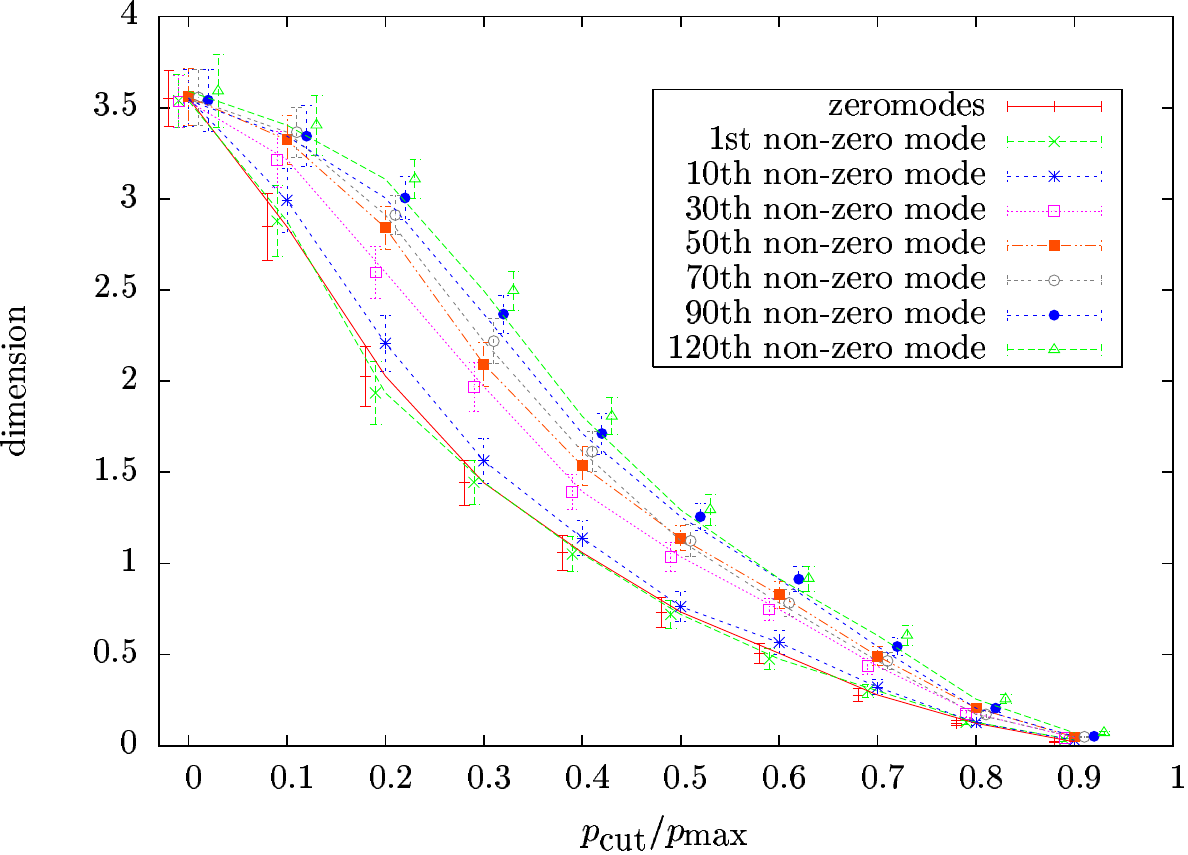}\\
\end{tabular}
\vspace{-0.2cm}
\caption{Cluster analysis of selected eigenmodes, averaged over the ensemble mentioned
in the text, as function of the cut-off $p_{\rm cut}/p_{\rm max}$ for the scalar density. 
Left: number of clusters in the mode; right: effective dimension $d^{*}$ of the mode.}
\label{fig:clustermodes}
\end{figure}

Fig.~\ref{fig:clustermodes} shows for selected modes in an ensemble of 170 lattices 
$16^3\times32$ at $\beta=8.45$ the number of clusters ({\it i.e.} separate maxima) 
on the left and the effective dimension $d^{*}$ of the clusters on the right as 
function of the cut-off $p_{\rm cut}$. The dimension was estimated by the random 
walker method. Percolation, that is not shown here, 
sets in between $p_{\rm cut}/p_{\rm max}<0.1$, 
{\it i.e.} rather low for the zero modes, and $p_{\rm cut}/p_{\rm max}=0.3$ for the 
120-th modes.

We conclude that all modes, including the zero modes, percolate at sufficiently low 
density. If cut at an average level of density, 5 to 20 different peaks are discernible, 
depending on the mode. 
The zero modes, before finally percolating, too, do not exceed a dimension $d^{*}=2$ 
as long as $p_{\rm cut}/p_{\rm max}>0.2$ .   

\section{Localization of topological charge} 
\begin{figure}[b]
\centering
\begin{tabular}{cc}
\includegraphics[height=4.5cm,width=7.1cm]{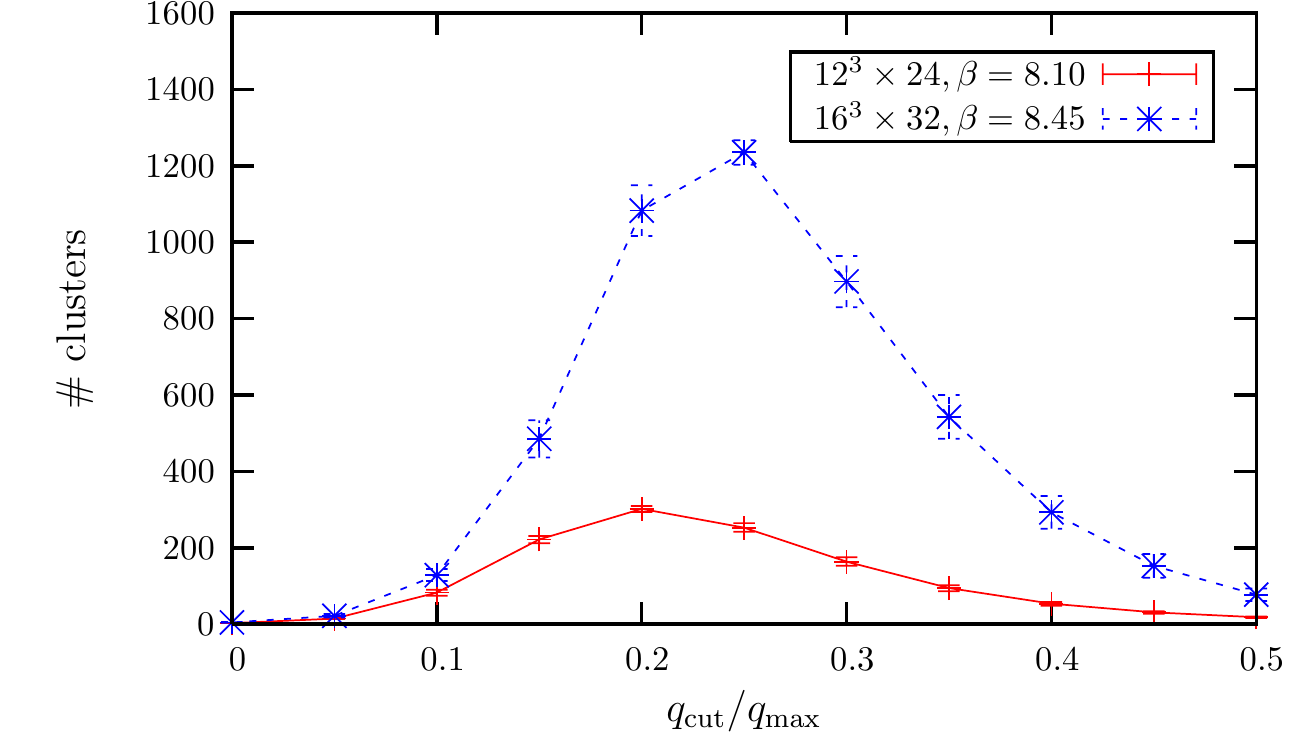}&
\includegraphics[height=4.5cm,width=7.1cm]{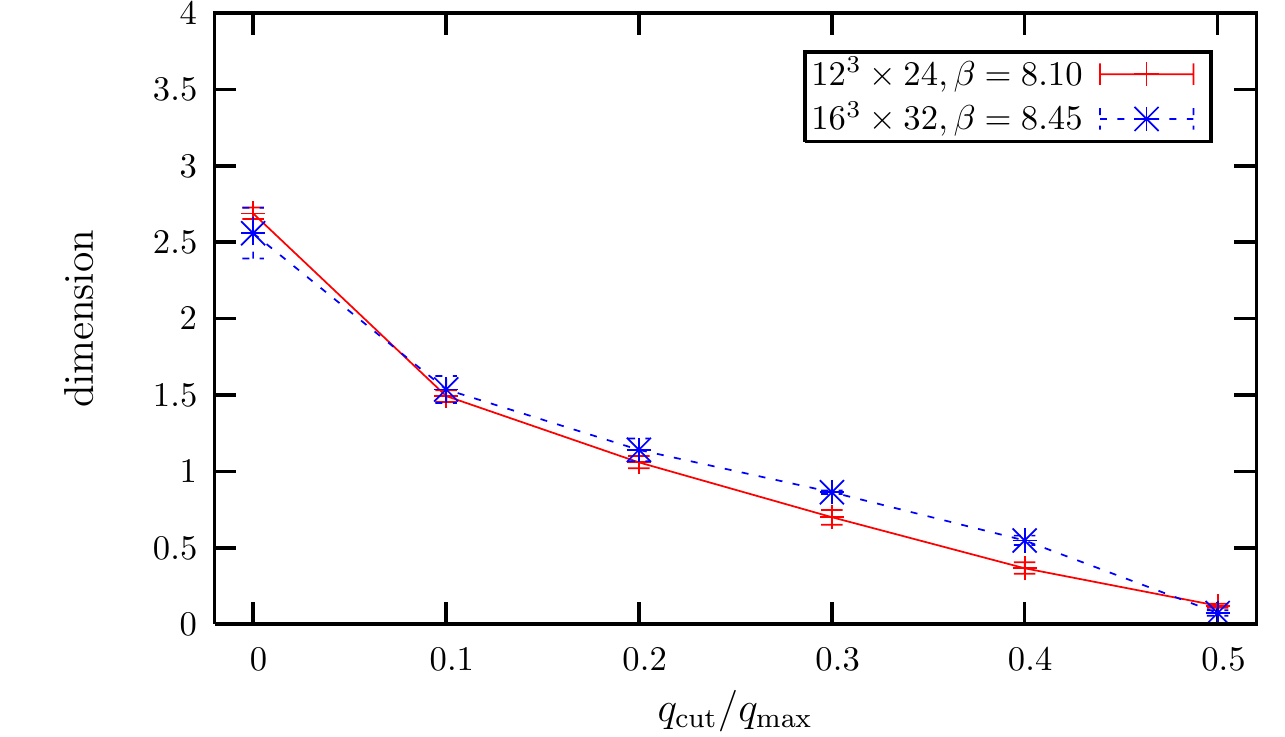}\\
\end{tabular}
\vspace{-0.2cm}
\caption{Cluster analysis of the all-scale topological density
for the $16^3\times32$ lattice at $\beta=8.45$ and the
$12^3\times24$ lattice at $\beta=8.1$ as function of the cut-off 
$q_{\rm cut}/q_{\rm max}$. Left: number of clusters; right: effective 
dimension of the clusters from the random walker method.} 
\label{fig:clusterfulltop}
\end{figure}

\begin{figure}[t]
\centering
\begin{tabular}{cc}
\includegraphics[height=4.5cm,width=7.1cm]{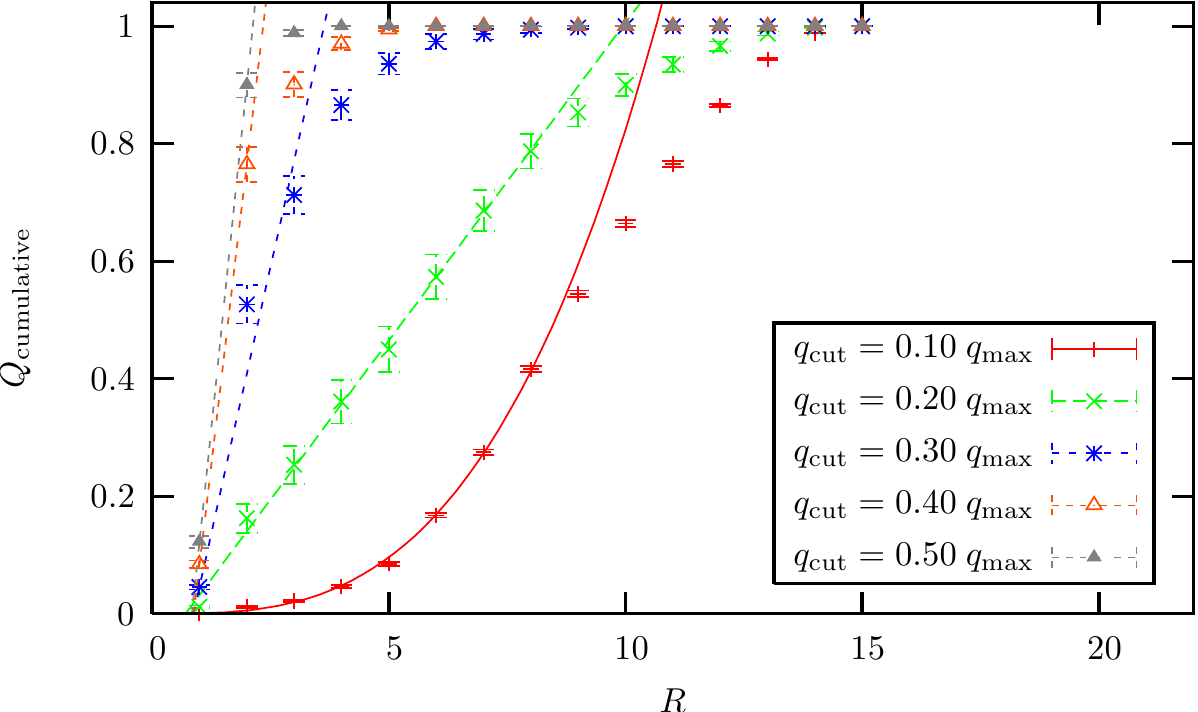}&
\includegraphics[height=4.5cm,width=7.1cm]{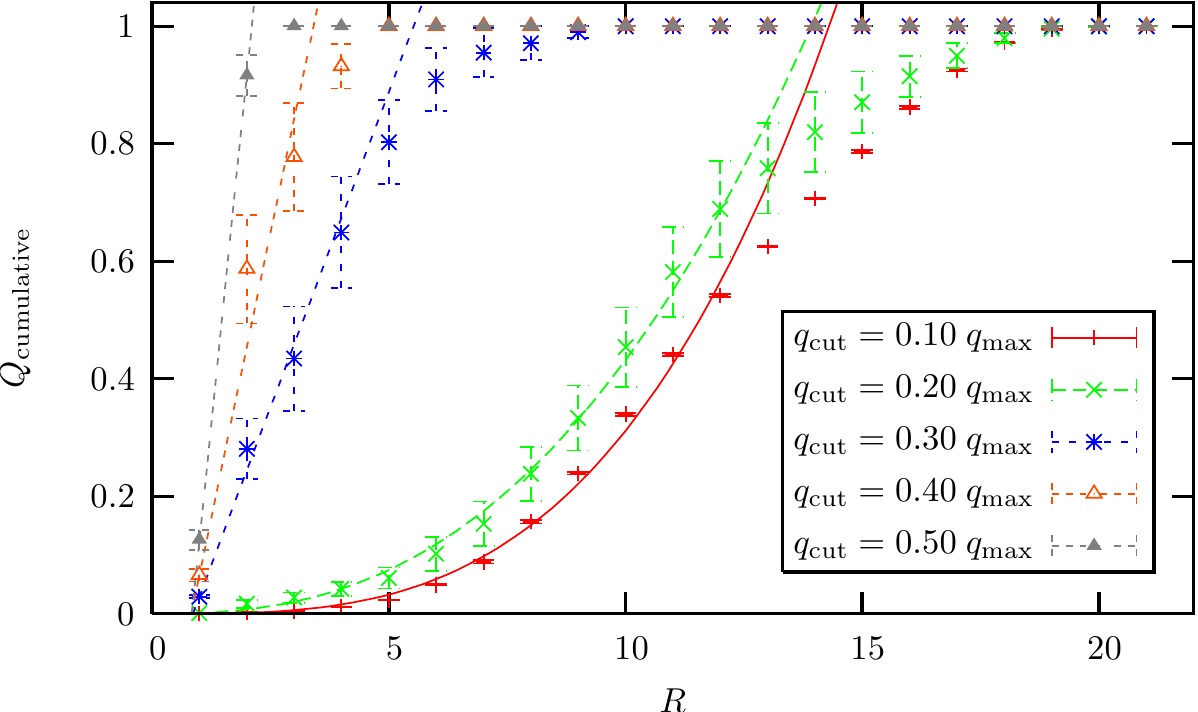}\\
\end{tabular}
\vspace{-0.2cm}
\caption{The ratio $Q_{\rm cumulative}(R)$ of the charge covered by the sphere
to the full cluster charge for the largest cluster is shown as function of $R$ for various
cut-off values, for the $12^3\times24$ lattice at $\beta=8.1$ (left) and the 
$16^3\times32$ lattice at $\beta=8.45$ (right). Note the abrupt change at 
$q_{\rm cut}/q_{\rm max}=0.2$ (left) and 0.25 (right).}
\label{fig:covering}
\end{figure}

\vspace{-0.2cm}
For a Dirac operator satisfying the Ginsparg-Wilson relation, the topological
charge density can be expressed as a local trace (with the massless overlap 
operator $D(0)$)
\begin{equation}
q(x) = - {\rm tr} \left[\gamma_5 \left(1 - \frac{a}{2} D(0;x,x) \right) \right] \; ,
\quad Q = \sum_x q(x) 
\end{equation}
over color and spinor indices. The UV filtered densities are obtained by casting 
this into a spectral sum and using a mode-truncation  $|\lambda| < \lambda_{\rm cut}$. 
Without truncation, we have evaluated the density only for a small subset of two of our
ensembles (53 configurations $12^3\times24$ for $\beta=8.1$ and 5 configurations 
$16^3\times32$ for $\beta=8.45$) representing almost equal volume. 
Fig.~\ref{fig:clusterfulltop} shows, analogously to the last figure, the cluster composition
and dimension of the {\it unfiltered} topological density as function of $q_{\rm cut}$ 
(meant as a cut-off for $|q(x)|$) for the two lattices.
The left plot shows that for $q_{\rm cut}/q_{\rm max} > 0.5$ only few isolated 
spikes are detected. They do not have a sufficient  extension that a power-law decay 
of $P(0,t)$ could be determined (``0-dimensional''). The right plot shows that for 
$q_{\rm cut}/q_{\rm max}  < 0.5$ the dimension starts growing to $d^{*} = 2.5$. At the 
maxima~\footnote{With $a \to 0$, the maximal number of clusters 
rises strongly.} of multiplicity, at $q_{\rm cut}/q_{\rm max}=0.2$ (for the coarse) 
and 0.25 (for the fine lattice), we find the same $d^{*} = 1$. Close to these maxima 
the covering sphere method also finds a change shown in Fig.~\ref{fig:covering}, 
signaling the onset of percolation. Below $q_{\rm cut}/q_{\rm max} < 0.05$ the 
cluster composition goes over to two sign-coherent global clusters of charge $Q_{+}$ and 
$Q_{-}$ which fill the volume and build the total charge $Q$ of the configuration. 
A distance between clusters, $C,C'$, can be defined as 
$\Delta_{C,C'} = \max_{x \in C} \left( \min_{y \in C'} |x-y| \right)$.
When the percolation is complete, the two remaining clusters have $\Delta \approx 2 a$, 
{\it i.e.} are closely intertwining each other.  

\section{Lower dimensional objects and confining vacuum defects}
\vspace{-0.2cm}
Here we shall offer an explanation for the even lower-dimensional local clusters 
of the unfiltered topological charge inside the global clusters (with $d^{*} \approx 2.5$).
For $1 < d^{*} < 2$, monopoles and vortices are good candidates to cause the
localization of charge. These are the two types of {\it confining defects} which are 
intimately connected~\cite{Boyko}. If one sort is removed, the other one disappears 
together with the topological charge (all zero modes) and the non-zero modes close to 
$\lambda=0$~\cite{Bornyakov}.
To demonstrate the correlation we have used the Indirect Maximal Center Gauge 
(IMCG)~\cite{IMCG}. In a first step the Maximally Abelian Gauge (MAG) is accomplished, 
followed by Abelian projection:
each MAG-fixed ${}^gU_{\rm link} \in SU(3)$ is replaced by the closest diagonal matrix 
$D_{\rm link} \in SU(3)$.
The norm $||{}^gU_{\rm link} - D_{\rm link}||$ is called non-Abelianicity. 
The monopole worldlines are located on cubes ({\it i.e.} links of the dual lattice) 
where the Abelian magnetic charges $(m^{(1)}_c,m^{(2)}_c,m^{{3}}_c)$ 
($\sum_k m^{(k)}_c=0$) are not all vanishing. 
In a second step, within residual Abelian gauge transformations $h \in U(1)^2$, 
we find the Maximal Center Gauge (MCG) which
brings ${}^hD_{\rm link}$ as close as possible to multiples of unity, 
$z_{\rm link} \times {\rm diag}(1,1,1)$ with 
$z_{\rm link} \in Z(3)$ being the links after center projection.
Center plaquettes, for which $p = \Pi_{{\rm link} \in \partial p}~z_{\rm link} \ne 1$, 
mark the presence of a vortex that is geometrically located on the 
dual (``vortex'') plaquette 
${}^{*}p$. A peculiarity of $SU(3)$ compared to $SU(2)$ is vortex splitting.

\begin{figure}[t]
\centering
\begin{tabular}{cc}
\includegraphics[height=4.5cm]{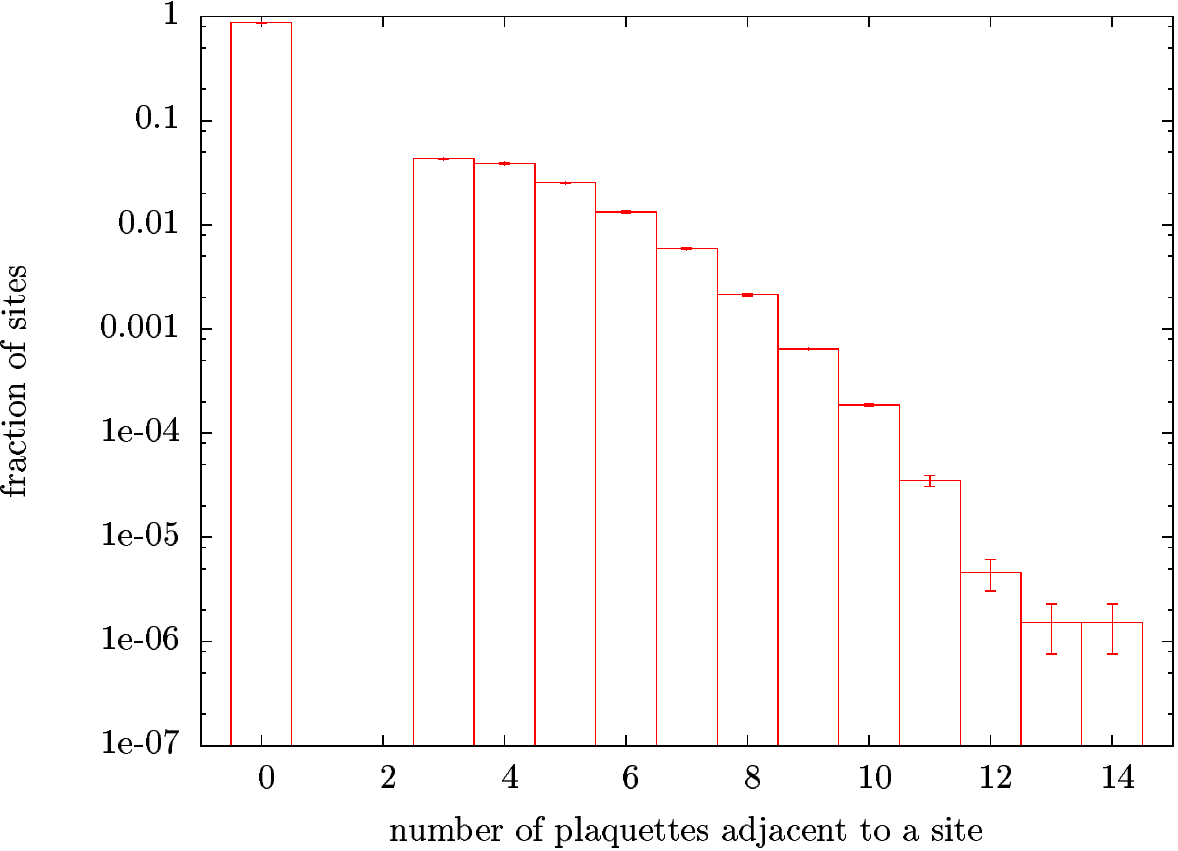}&
\hspace{1cm}
\includegraphics[height=4.5cm]{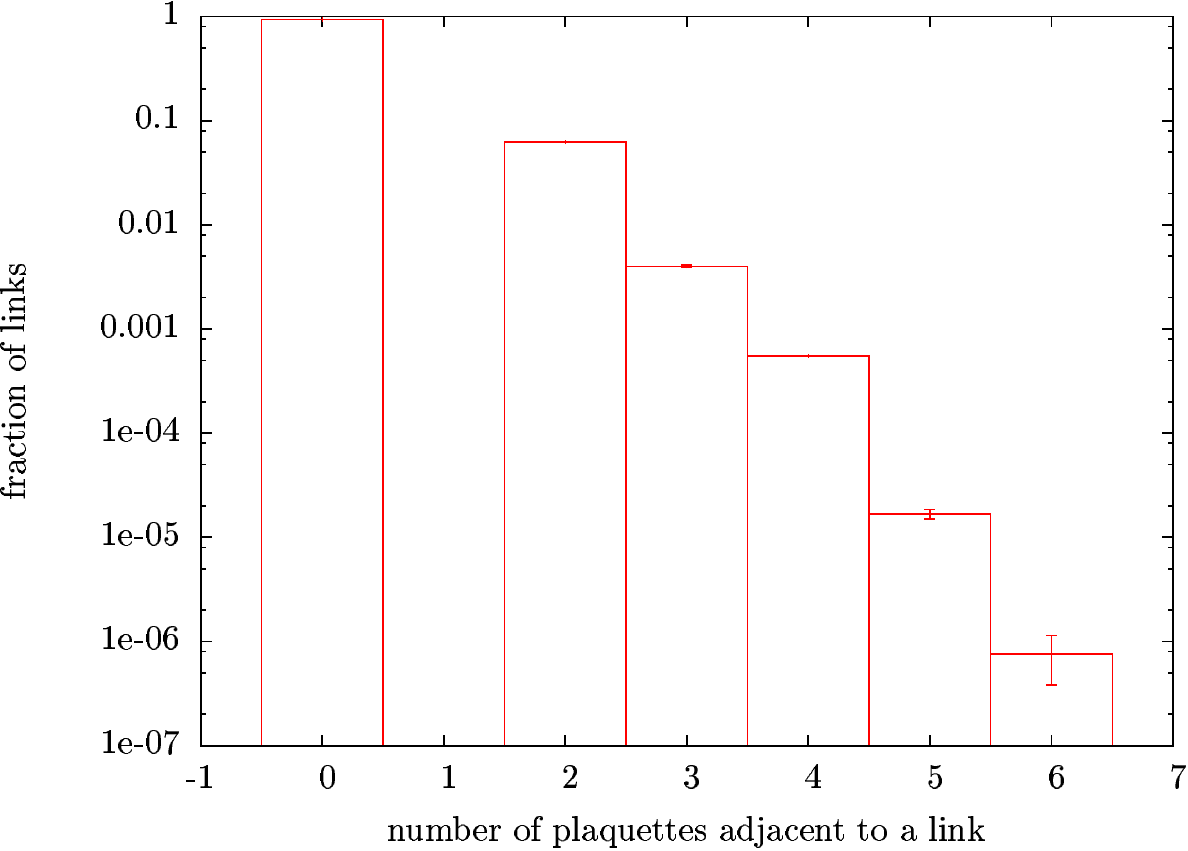}\\ 
\end{tabular}
\vspace{-0.2cm}
\caption{The connectivity of the vortex structure.
Left: histogram of dual sites w.r.t. the number $n$ of adjacent vortex plaquettes; 
Right: histogram for dual links w.r.t. the number $n$ of adjacent vortex plaquettes.}
\label{fig:incidence}
\end{figure}

Density and connectivity describe the ``vortex matter'' corresponding to some 
given gauge field ensemble. For $\beta=8.45$ we find the probability for a dual site
to be adjacent to $n$ vortex plaquettes as shown by the histogram in the left of 
Fig.~\ref{fig:incidence}. Here 87 \% of sites belong to the bulk ($n=0$), 
4.3 \% are adjacent to 3 plaquettes (corner), 3.8 \% to 4 plaquettes (planar vortex) etc.
The probability for a dual link to be adjacent to $n$ vortex plaquettes is shown in 
the histogram on the right of Fig.~\ref{fig:incidence}. This is an important input 
for the construction of a realistic effective vortex model~\cite{Quandt} for confinement.
Thus 93 \% of the links belong to the bulk ($n=0$), 6.25 \% are adjacent to 
$n=2$ plaquettes, 0.4 \% to $n=3$ plaquettes (branching) etc. 

\begin{figure}[b]
\centering
\begin{tabular}{cc}
\includegraphics[height=4.5cm]{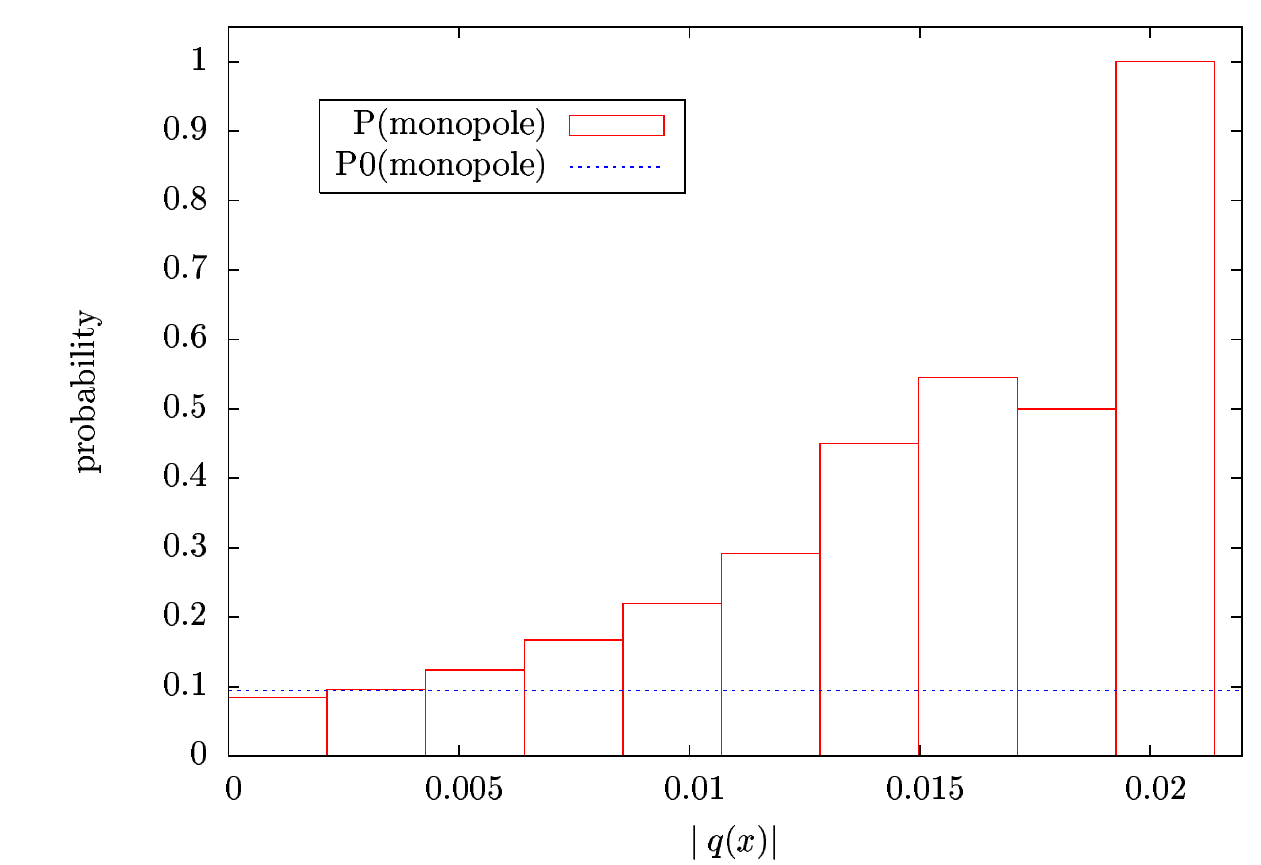}&
\hspace{1cm}
\includegraphics[height=4.5cm]{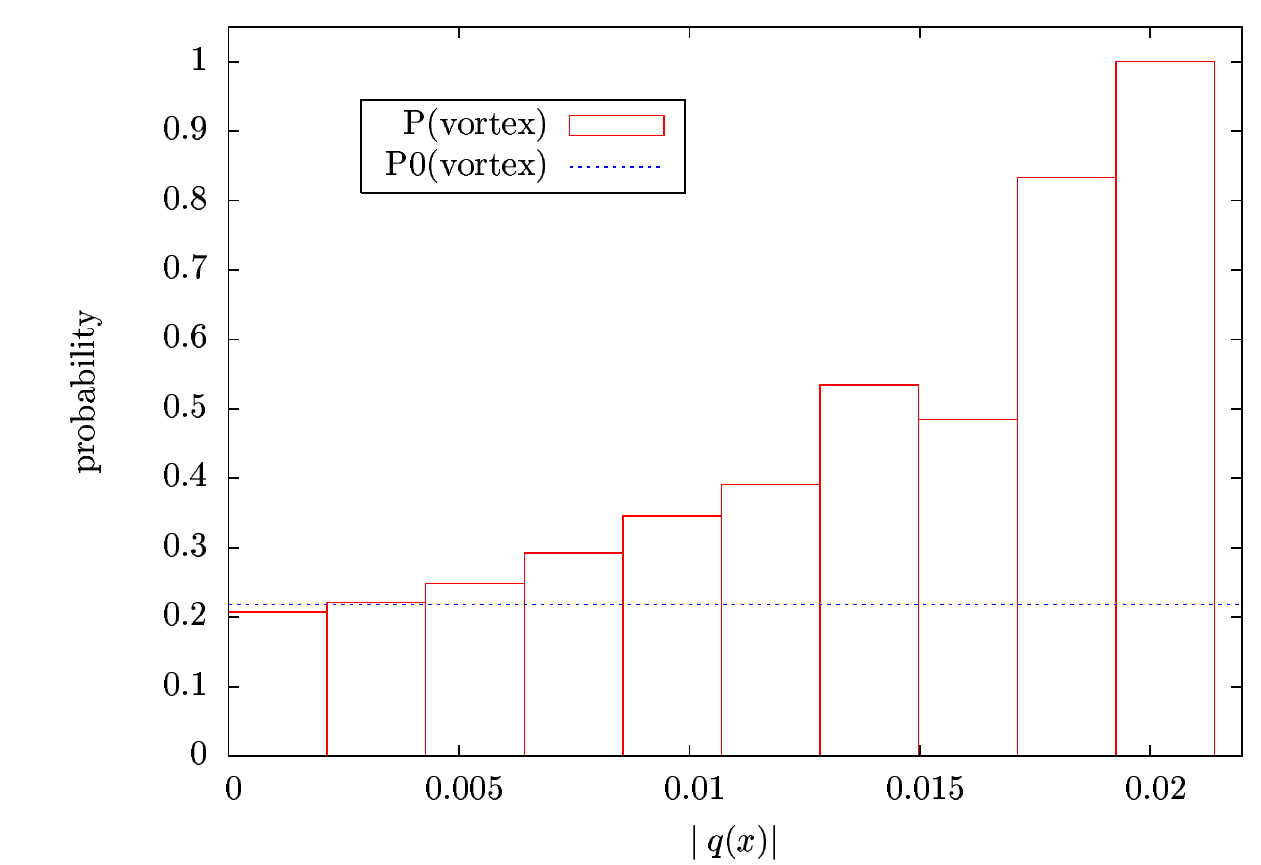}\\ 
\end{tabular}
\vspace{-0.2cm}
\caption{The probability P for a site of the original lattice to be adjacent (closest) 
to a monopole (left) and to a vortex (right) depending on the unfiltered topological 
density. The horizontal blue lines show the 
{\it a priori} probability P0 for a site to be 
close to a monopole or vortex. The histogram covers the interval from $0$ 
to $q_{\rm max}$.}
\label{fig:probabilities}
\end{figure}

Close to the monopoles, the non-Abelianicity and the modulus of the topological 
density $|q(x)|$ show an excess above their bulk averages. Thus they are positively 
correlated.
Fig.~\ref{fig:probabilities} illustrates the enhanced probability to find a site
of the original lattice close to a monopole and/or vortex if for the (unfiltered)
topological density at the site $|q(x)| > 0.2~q_{\rm max}$ is fulfilled. 
This proves that the confining defects are the preferred location for topological 
charge. 

\section{Summary}
We have summarized the localization of eigenmodes and of the unfiltered all-scale
topological density $q(x)$ provided by our recent investigation of the vacuum structure 
based on the overlap operator~\cite{Ilgenfritz1}.
In addition, we have presented first results relating the low dimensionality of $q(x)$ 
at densities above the percolation threshold to a local correlation with monopoles 
and vortices detected in the course of IMCG. More results and 
corresponding observations concerning the lowest modes will be published 
elsewhere~\cite{elsewhere}.

\section*{Acknowledgements}

The numerical overlap calculations have been performed on the IBM p690 at HLRN
(Berlin) and NIC (J\"ulich), as well as on the PC farms at DESY Zeuthen and
LRZ Munich. We thank these institutions for support. Part of this work is 
supported by DFG under contract FOR 465 
(Forschergruppe Gitter-Hadronen Ph\"anomenologie).


\end{document}